\documentclass[aps,prd,floatfix,preprint,nofootinbib]{revtex4-1}
\usepackage{graphicx}
\usepackage{epic}
\usepackage{eepic}
\usepackage{latexsym}
\usepackage{amssymb,amsmath}

\newcommand{\eq}[1]{(\ref{#1})}
\newcommand{\be}{\begin{equation}}
\newcommand{\ee}{\end{equation}}
\newcommand{\bea}{\begin{eqnarray}}
\newcommand{\eea}{\end{eqnarray}}

\newcommand{\hs}[1]{\hspace{#1 mm}}

\newcommand{\lf}{\left<}
\newcommand{\rg}{\right>}
\newcommand{\vv}{|0\right>}
\newcommand{\vc}{\left<0|}
\newcommand{\dd}{\delta_\rho}
\newcommand{\cb}{\bar{\chi}}
\newcommand{\I}{(f)}

\def\cc{\gamma}
\def\C{\Gamma}
\def\d{\delta}

\def\e{\epsilon}

\def\f{\phi}
\def\fr{\frac}
\def\F{\Phi}
\def\vf{\varphi}

\def\l{\lambda}

\def\m{\mu}
\def\n{\nu}

\def\r{\rho}

\def\th{\theta}

\def\z{\zeta}
\def\O{\Omega}

\def\del{\partial}

\let\la=\label
\let\bm=\bibitem
\def\nn{\nonumber}

\begin{document}

\title{Issues about Cosmological Ward Identities} 

\author{Ali Kaya}
\email[]{ali.kaya@boun.edu.tr}
\affiliation{Bo\~{g}azi\c{c}i University, Department of Physics, 34342, Bebek, \.{I}stanbul, Turkey}

\date{\today}
\begin{abstract}
In this paper we first discuss how a Noether current corresponding to a gauge or a global symmetry can locally be introduced in a path integral irrespective of the boundary conditions defining the theory. We then consider quantization of gravity plus minimally coupled scalar field system in the phase space path integral approach. The complete gauge fixed action including the Faddeev-Popov determinant is obtained in the so called $\z$-gauge. It turns out that in this formalism while the dilatation survives as the residual symmetry of the gauge fixed action, other diffeomorphisms which require field dependent corrections fail to be so. The full Noether current for the dilatation is determined and the spatial boundary conditions that yield a finite and conserved charge are determined. The charge is shown to be expressible as a surface integral at infinity and the corresponding Ward identity gives the standard consistency relation of cosmological perturbations. 
 
\end{abstract}

\maketitle

\section{Introduction}

For now, cosmology seems to be the only testing ground for quantum gravity. Although the exact theory, which should presumably resolve the big-bang singularity, is not known one may nevertheless consider small fluctuations around a classical background and remarkably the results obtained in the cosmological perturbation theory are consistent with the observations. In recent years the linearized theory has been developed to include interactions \cite{mal,w1,w2,w3} but the approach is still perturbative and the issues about renormalization, which are intricate when gravity is involved, are  mainly overlooked. Not surprisingly, symmetry considerations provide interesting non-perturbative information about cosmological perturbations; Maldacena's  consistency relation is being an example \cite{mal}

In the presence of a gauge symmetry, the quantum theory requires a viable gauge fixing which breaks the local invariance, but even in that case some residual symmetries may remain in the theory. For cosmological perturbations in the scalar slow-roll inflationary models, in the so called $\z$-gauge where the time slices are chosen to kill the scalar field fluctuations, an infinite set of residual symmetries have been shown to exist in the literature \cite{in1,in2}. These are spatial diffeomorphisms  that preserve the transversality of the tensor mode, however all but the dilatation symmetry require field dependent corrections to the gauge parameter. Naively, each residual diffeomorphism gives a Ward identity involving cosmological correlation functions (see e.g. \cite{s0,s1,s2,s3,s4,s5,s6,s7,s8}). 

It is somehow surprising that the residual symmetries of cosmological perturbations are identified in the literature without anyway discussing the boundary conditions. As we will see (and as it is likely well known by many experts) a Noether current corresponding to a local or a global symmetry always exists irrespective of the boundary conditions and possible boundary terms in the action. Namely, when an action is invariant under a symmetry transformation up to surface terms, which may or may not vanish depending on the boundary conditions, a conserved Noether current can always be found. Moreover, the charge density of this current can be shown to generate the symmetry transformation in the quantum  theory. This is a purely local result which must follow from the field equations and the canonical commutation relations. On the other hand, proper boundary conditions are needed to get a well defined and conserved charge whose existence should lead nontrivial information. 

Having inflation in mind, in this paper we consider the standard Einstein gravity coupled to a self interacting real scalar field. Since the ADM decomposition is conveniently used for cosmological perturbations, we utilize the (formal) phase space path integral quantization of the system. In practice, the fields must be expanded around a classical cosmological background but we try to keep the discussion more general by not referring to this perturbative expansion as long as possible. The diffeomorphism invariance is fixed by imposing the $\z$-gauge and the corresponding Faddeev-Popov determinant is calculated.  As a result, the phase space path integral involves the Einstein-Hilbert action in the Hamiltonian form, the gauge fixing terms and a ghost action related to the Faddeev-Popov determinant. We show that the complete action is invariant under the dilatation, which becomes the residual symmetry of the quantum theory\footnote{In stating this,  we assume that the formal  nonperturbative phase space path integral exists and we simply ignore issues related to renormalization, seeing them as artifacts of the perturbation theory.} with suitable boundary conditions, and we determine the corresponding Noether current. On the other hand, the invariance of the theory under the residual diffeomorphisms that require field dependent parameters is dubious in this formalism since the corresponding map of fields is not canonical and thus the integration measure picks up a nontrivial Jacobian. 

Most of the time, the Noether charge associated to a local symmetry can be expressed as a surface integral at infinity (see e.g. \cite{c1,c2,c3,c4}). In our problem, this is fairly evident in the covariant theory and we explicitly show it to be true when the fields are expanded around a cosmological background. We determine the spatial boundary conditions which are required for a finite and conserved charge. Together with the prescribed boundary conditions, the theory becomes free near the spatial infinity. This allows us to relate the Noether charge to a zero mode and the corresponding Ward identity gives the standard consistency relation of cosmological perturbations. 

\section{The Noether Current}

In this section, we first review a few salient features of the Noether current in the path integral quantization and later study the gravity plus scalar field system. Our approach is mostly motivated by and very similar to \cite{n}, but there are also some notable differences. Consider the following elementary in-out path integral of a free massless scalar field in the flat space with the standard action $S=-1/2\int\del_\m\f\del^\m\f$ (since we mainly discuss the in-in path integrals below, the time ordering of operators will always be indicated explicitly)
\be\label{1}
\int  D\f\,e^{iS}\f(x_1^\m). . . \f(x_n^\m)=\vc T\f(x_1^\m). . . \f(x_n^\m)\vv.
\ee
Clearly the action is invariant under a constant (infinitesimal) shift $\d_S\f\equiv c$. Assuming naively that the path integral measure is also invariant under this shift leads to the bizarre conclusion that all Green functions must vanish. Indeed, this assumption is incorrect since the path integral is over all fields which vanish at infinity allowing both integration by parts in the action and Fourier transformation of the fields. Unfortunately, the innocent looking constant shift does not respect this boundary condition. To revive the shift symmetry in the path integral one may try to apply the transformation only in a local region. For that consider the following deformation of the symmetry 
\be\label{2}
\dd\f\equiv\r(x)\d_S\f=\r(x)c,
\ee
where $\r$ is an arbitrary function of compact support in the space-time. Applying the deformed transformation \eq{2} to the path integral \eq{1}, one may find
\be
\int D\f\, e^{iS}\left[\dd\left(\f_1. . . \f_n\right)+i\int d^4 x \,\r(x)\,[\del^2\f(x)](\f_1 . . . \f_n)\right]=0,\label{3}
\ee
where $\f_1=\f(x_1^\m)$ and so on. Note that $\dd S=-\int \del_\m\f\del^\m\r=\int \r\del^2\f$ since $\r$ has compact support. The path integral measure is now invariant under \eq{2}, which respects whatever boundary conditions one has in the theory. Since \eq{3} is true for any $\r$, one concludes 
\be
-i\del^2\vc T\f_1. . . \f_n\f(x)\vv=\d^4(x-x_1)\vc T\f_2. . . \f_n\f(x)\vv+. . . +\d^4(x-x_n)\vc T \f_1. . .\f_{n-1}\f(x)\vv,\label{4}
\ee
which can easily be verified by Wick's theorem.   

The above example is simple but it highlights the crux of the matter. Consider a theory governed by an action $S[\F^I]$, where the fields are collectively denoted by $\F^I$. Assume that the action is invariant up to possible surface terms under a global or a local infinitesimal symmetry transformation $\d_S\F^I(x)$. Define the locally deformed transformation by multiplying with an arbitrary function of compact support $\r$ as 
\be\label{5}
\dd\F^I(x)\equiv \r(x)[\d_S\F^I(x)].
\ee
It is easy to see by integration by parts that the variation of the action under \eq{5} can be written as
\be\label{6}
\dd S=-\int d^4x\,\r(x)\left[\del_\m J^\m\right],
\ee
where $J^\m$ can be identified as the Noether current. Eq. \eq{6} follows from the fact that $\r=1$ is the original symmetry transformation which leaves the action invariant up to the surface terms. When the field equations hold, \eq{6} should vanish for arbitrary $\r$ that gives $\del_\m J^\m=0$. Note that the Noether current defined in this way is not affected by the boundary conditions or by possible boundary terms in the action. The current is not unique either, because 
\be\la{fr}
J^\m\to J^\m+\del_\n K^{\m\n}
\ee
with arbitrary $K^{\m\n}=K^{[\m\n]}$ still yields \eq{6}.

Consider now the following in-in path integral giving the expectation value of an operator $O(t)$ in a (vacuum) state
\be
\int [D\F^I_+][ D\F^I_-][D\F^I_*]\, e^{iS_+-S_-}\, O_+(t)\equiv \lf O(t)\rg,\label{7}
\ee
where $O(t)=\F^{I_1}(t,\vec{x}_1). . .\F^{I_n}(t,\vec{x}_n)$ and for notational simplicity the initial state wave-functionals are omitted.\footnote{We will never apply a transformation to the path integral extending through the initial time $t_i$, therefore the state wave-functionals will always remain intact.} Having cosmological applications in mind, in \eq{7} we consider a composite operator defined at a single time but the following argument can easily be generalized to other cases.  We take $t<t_*$ to avoid field insertions at the fixed return time $t_*$. The above path integral is over all fields $\F_+^I$ and $\F_-^I$ satisfying $\F_*^I=\F_+^I(t_*)=\F_-^I(t_*)$ and $[D\F_*^I]$ denotes the spatial integration measure of the field variables $\F_*^I$ at the constant $t_*$ slice. As shown in \cite{ali}, it is possible to carry out the path integral over $\F_*^I$ at $t_*$, which simply imposes the extra condition  $\dot{\F}_+^I(t_*)=\dot{\F}_-^I(t_*)$ for $\F^I_\pm$ fields. Namely, the in-in path integral can be carried out either by integrating  $\F_+^I$, $\F_-^I$ and $\F_*^I$ obeying $\F_*^I=\F_+^I(t_*)=\F_-^I(t_*)$; or by integrating over the fields $\F_+^I$ and $\F_-^I$ satisfying $\F_+^I(t_*)=\F_-^I(t_*)$ and $\dot{\F}_+^I(t_*)=\dot{\F}_-^I(t_*)$. Inevitably, these integrals also require certain boundary conditions at spatial infinity, which are irrelevant at the moment. 

Let us see how the infinitesimal transformation \eq{5} applies to the in-in path integral \eq{7}. Introduce two independent functions of compact supports $\r_+$ and $\r_-$ obeying $\r_+(t_*)=\r_-(t_*)=0$ and define
\be
\d_+\F_+^I=\r_+[\d_S\F_+^I],\hs{10}\d_-\F_-^I=\r_-[\d_S\F_-^I].\label{8}
\ee
We prefer not to extend the deformed transformations through the return time $t_*$ to avoid boundary effects related to the $[D\F_*^I]$ integral (we also take  $\r_+(t_i)=\r_-(t_i)=0$, where $t_i$ is the initial time). It is natural to assume that the in-in path integral measure is invariant under \eq{8}, which only restricts the symmetry transformation in a local region. Then,  by applying \eq{8} to \eq{7} with $\r_+=0$ one may easily see that
\be\label{9}
\lf O(t)[\del_\m J^\m]\rg=0.
\ee
On the other hand, replacing $O_+$ in \eq{7} by $O_-$ and setting this time $\r_-=0$ gives
\be\label{10}
\lf [\del_\m J^\m] O(t)\rg=0.
\ee
These two equations show that the Noether current is also conserved in the quantum theory. Choosing now $\r_-=0$ and $\r_+\not=0$ in \eq{7} implies
\be\label{11}
i\del_\m\lf T O(t)J^\m(x)\rg=\d^4(x-x_1)\lf [\d_S\F^{I_1}(t,\vec{x}_1)]. . .\F^{I_n}\rg+. . . +\d^4(x-x_n)\lf \F^{I_1}. . . [\d_S\F^{I_n}(t,\vec{x}_n)]\rg.
\ee
Finally, using \eq{9} and the explicit definition of the time ordering in \eq{11} yields
\be\label{12}
i\lf\left[J^0(t,\vec{x}),O(t)\right]\rg=\d^3(\vec{x}-\vec{x}_1)\lf\ [\d_S\F^{I_1}]. . .\F^{I_n}\rg+. . .+ \d^3(\vec{x}-\vec{x}_n)\lf \F^{I_1}. . . [\d_S\F^{I_n}]\rg.
\ee
This last equation shows that the Noether charge density is the generator of the symmetry transformation in the quantum theory. In getting these identities there is no need to refer to the boundary conditions since the deformed transformations \eq{8} do not extend to infinity. Therefore, these should follow from purely local physics. More explicitly, the current conservation $\del_\m J^\m=0$ is expected to hold by field equations and the generator equation \eq{12} presumably  involves the canonical commutation relations. 

The above discussion clarifies the basic role of the Noether current in the path integral quantization and this is as far as one may continue without paying attention to the boundary conditions. Obviously, proper boundary conditions are required for the Noether charge $Q(t)=\int d^3 x J^0(t,\vec{x})$ to be well defined and further for it to be conserved $dQ/dt=0$.  

After these general considerations, let us now focus on our main interest, i.e. the gravity plus minimally coupled scalar field system. We take the metric in the ADM form 
\be\label{13}
ds^2=-N^2dt^2+h_{ij}(dx^i+N^i dt)(dx^j+N^j dt),
\ee
and study the theory in the Hamiltonian formulation. The Einstein-Hilbert action can be written as
\be\label{14}
S_{EH}=\int d^4x \,{\cal L}=\int d^4x\left[\Pi^{ij}\dot{h}_{ij}+P_\f\dot{\f}-N\F-N^i\F_i\right]
\ee
where the dot denotes the time derivative, 
\bea
&&\F=\fr{1}{\sqrt{h}}\left[\Pi^{ij}\Pi_{ij}-\fr12\Pi^2\right]+\fr{1}{2\sqrt{h}}P_\f^2+\sqrt{h}\left[V(\f)+\fr12 h^{ij}\del_i\f\del_j\f-R^{(3)}\right],\nn\\
&&\F_i=-2\sqrt{h}D_j\left[\fr{1}{\sqrt{h}}\Pi^{j}{}_i\right]+P_\f\del_i\f,\label{15}
\eea
$D_i$ is the covariant derivative and $R^{(3)}$ is the Ricci scalar of $h_{ij}$, $h=\det(h_{ij})$ and $\Pi=\Pi^{ij}h_{ij}$. In this section all index manipulations are carried out by the spatial metric $h_{ij}$. In the classical theory the canonical pairs obey the following Poisson brackets
\bea
&&\left\{h_{ij}(\vec{x}),\Pi^{rs}(\vec{y}\right\}=\fr12\left(\d^r_i\d^s_j+\d^r_j\d^s_i\right)\d^3(\vec{x}-\vec{y}),\nn\\
&&\left\{\f(\vec{x}),P_\f(\vec{y}\right\}=\d^3(\vec{x}-\vec{y}),\label{16}
\eea
and the lapse $N$ and the shift $N^i$ are Lagrange multipliers enforcing the Hamiltonian and the momentum constraints; $\F=0$ and $\F_i=0$. 

In the Lagrangian formulation, the theory is invariant under the full diffeomorphism group generated by the vector fields $k^\m=(k^0,k^i)$ due to the underlying geometric structure. In the next section we will fix the time reparametrizations and thus in the following we only consider coordinate changes with $k^0=0$.  On the other hand, in the Hamiltonian formulation a time dependent spatial diffeomorphism generated by $k^\m=(0,k^i(t,\vec{x}))$ acts like a time dependent canonical transformation which generates extra terms in the action. Although it is possible to deal with this issue, it is enough for our purposes to focus on time independent spatial maps generated by a 3-vector $k^i(\vec{x})$ whose action on the fields become
\bea
&&\d_S h_{ij}=k^r\del_r h_{ij}+h_{rj}\del_ik^r+h_{ir}\del_jk^r=D_ik_j+D_jk_i,\nn\\
&&\d_S\Pi^{ij}=k^r\del_r\Pi^{ij}-\Pi^{rj}\del_rk^i-\Pi^{ri}\del_rk^j+\Pi^{ij}\del_rk^r\equiv {\cal L}_{\vec{k}}\Pi^{ij},\nn\\
&&\d_S\f=k^i\del_i\f,\label{17}\\
&&\d_SP_\f=k^i\del_iP_\f+P_\f\del_rk^r,\nn\\
&&\d_S N=k^i\del_i N,\nn\\
&&\d_S N^i=k^r\del_r N^i-N^r\del_rk^i.\nn
\eea
It is worth to emphasize that since $k^i$ is chosen to be time independent, all fields and their time derivatives transform like 3-dimensional tensors (or tensor densities of weight one), which live in the tangent space of constant time slices. The variation of the action \eq{14} under \eq{17} is then given by a total surface term
\be\label{18}
\d_S S_{EH}=\int d^4 x\, \del_i\,(k^i{\cal L}).
\ee
This key geometric structure will be very useful for our subsequent considerations.  

The Noether current corresponding to \eq{17} can be calculated by deforming the transformations\footnote{We would like to point out that while the transformation \eq{17} is a canonical map in the  phase space, the deformed transformation is {\it not} canonical. To see this, one can verify that the Poisson bracket structure is not invariant $G(x,y)\equiv\d_\r\left\{h_{ij}(x),\pi^{rs}(y)\right\}\not=0$. Since $G(x,y)$ is actually a distribution, one may calculate $\int d^3xd^3yf(x)g(y)G(x,y)$ for arbitrary compact functions $f(x)$ and $g(y)$. Using \eq{16}, we obtain $\int d^3xd^3yf(x)g(y)G(x,y)=-\fr12(\d^r_i\d^s_j+\d^s_i\d^r_j)\int d^3 x f(x)g(x)k^l(x)(\del_l\r)$, which does not vanish unless $\r$ is a constant. This issue was a reason for debate about IR divergences of massless fields in de Sitter space, see \cite{deb1,deb2,deb3}.} like
\be
\dd h_{ij}=\r[\d_S h_{ij}]=\r(D_ik_j+D_jk_i)\label{19}
\ee
where this modification should be applied to all equations in \eq{17}. A straightforward but lengthy calculation then gives
\be
\dd S_{EH}=-\int d^4 x \,\r\,[\del_\m\hat{J}^\m],\label{20}
\ee
where
\bea
&&\hat{J}^0=2\Pi^{ij}D_ik_j+P_\f k^i\del_i\f,\nn\\
&&\hat{J}^i=-k^i{\cal L}-N^iP_\f k^r\del_r\f-\sqrt{h}(D^i\f)k^r\del_r\f+2N^r{\cal L}_{\vec{k}}\Pi^{i}{}_r-2N^i\Pi^{rs}D_rk_s\label{21}\\
&&\hs{10}+2\sqrt{h}(D^iN)D_rk^r+2\sqrt{h}ND^rD_rk^i-2\sqrt{h}ND^iD_rk^r-2\sqrt{h}(D^rN)D^ik_r.\nn
\eea
As noted before, $\hat{J}^\m$  is not unique and one may obtain different Noether currents still satisfying \eq{20}. It is important to observe that
\be\label{22}
\hat{J}^0=k^i\F_i+2\del_i[\Pi^{ij}k_j],
\ee
and thus on shell when the momentum constraint $\F_i=0$ is satisfied, the Noether charge becomes a surface integral at spatial infinity. This observation will be crucial in getting the cosmological consistency relation from the Ward identity. 

\section{Gauge Fixing, Faddeev-Popov Determinant and the Residual Symmetry}

In this section we would like to apply the phase space path integral quantization of the gravity plus scalar field system. Since there is no known non-perturbative quantization procedure, one must actually study the fluctuations around a classical background and apply perturbation theory. Even in that case the renormalization of the theory is problematic since gravity is involved. Here, we simply ignore these issues and keep the quantization procedure formal. To have closed form expressions, we also would like to postpone the expansion around the classical background as long as possible. 

There are four primary constraints $\F=0$ and $\F_i=0$ that demand four gauge conditions. We utilize the so called $\z$-gauge as follows: The time reparametrizations can be fixed by imposing
\be\label{23}
G\equiv \f-\f_B(t)=0,
\ee
where $\f_B(t)$ can be identified as the background value of the scalar field that obeys $\dot{\f}_B\not=0$ and otherwise (for now) arbitrary. This condition completely breaks the diffeomorphism invariance with $k^0\not=0$. To fix spatial diffeomorphisms we introduce $\d_{ij}$ as a background metric, where the indices refer to the ADM coordinates introduced in \eq{13}. We define the trace-free graviton field as
\be\label{24}
\cc_{ij}\equiv h_{ij}-\fr13\d_{ij}\d^{mn}h_{mn}\equiv h_{ij}-\fr13\d_{ij}h_{mm}, 
\ee
which obeys $\d^{ij}\cc_{ij}\equiv\cc_{ii}=0$. Since there are now two metric tensors $h_{ij}$ and $\d_{ij}$, index manipulations should be done with care. We never raise or lower the indices with $\d_{ij}$ and for notational simplicity the summations involving $\d_{ij}$ will be denoted like $\d^{ij}h_{ij}=h_{ii}$. We impose 
\be\label{25}
G_i\equiv \del_j\cc_{ji}=0,
\ee
which completes the gauge fixing. 

In the path integral quantization, the gauge conditions \eq{23} and \eq{25} can be implemented by Lagrange multipliers $\l^\m=(\l,\l^i)$ via the gauge fixing action
\be\label{26}
S_{GF}=\int d^4 x\,(\l G+\l^iG_i),
\ee
which must be added to the Einstein Hilbert action \eq{14}. 

There are now four primary constraints $\F_\m=(\F,\F_i)$ and four gauge conditions $G_\m=(G,G_i)$. In the Hamiltonian formalism the corresponding Faddeev-Popov determinant is given by
\be\label{27}
M=\det\left\{\F_\m(t,\vec{x}),G_\n(t,\vec{y})\right\},
\ee
where the Poisson brackets must be found using \eq{16}. The functional determinant $M$ can be calculated by introducing anti-commuting ghost and anti-ghost fields, $\chi^\m=(\chi,\chi^i)$ and $\cb^\m=(\cb,\cb^i)$, and the Faddeev-Popov ghost action as
\be
S_{FP}=\int dt\, d^3x\,d^3y\,\cb^\m(t,\vec{x})\left\{\F_\m(t,\vec{x}),G_\n(t,\vec{y})\right\}\,\chi^\n(t,\vec{y}).\la{28}
\ee
After a relatively long but straightforward calculation we obtain
\bea
&&S_{FP}=\int d^4x\left[-\fr{1}{\sqrt{h}}\cb\chi P_\f-\cb^i\chi\del_i\f+\fr{1}{\sqrt{h}}(\cc_{ij}\Pi-2\tilde{\C}_{ij})\cb\del_j\chi^i+(\del_i\cc_{kj})\cb^i\del_k\chi^j\right.\nn\\
&&\left.\hs{13}-h_{ij}\del_k\cb^i\del_k\chi^j-h_{ik}\del_j\cb^i\del_k\chi^j+\fr23h_{ik}\del_k\cb^i\del_j\chi^j\right],\label{29}
\eea
where $\tilde{\C}_{ij}=\Pi_{ij}-\d_{ij}\Pi_{rr}/3$. 

The formal in-out phase space path integral quantization of the system involves the following integral over field variables obeying suitable boundary conditions 
\be\label{30}
DX\,e^{iS}
\ee
where
\be\la{31}
DX\equiv Dh_{ij}D\Pi^{kl}D\f DP_\f DN DN^m D\l^\r D\cb^\m D\chi^\n
\ee
and $S=S_{EH}+S_{GF}+S_{FP}$. While the integrals over $N$, $N^i$ and $\l^\m$ impose the constraints and the gauge conditions, the ghost integrals yield the corresponding Faddeev-Popov determinant. These eliminate the gauge degrees of freedom and select out the physical subspace in the unconstrained phase space. For the in-in case, the path integral includes $+$ and $-$  branches and a spatial path integral defined at the return time $t_*$ so that
\be\la{32}
DX_+DX_-DX_*e^{iS_+-iS_-}.
\ee
The full action is quadratic in the momenta $\Pi^{ij}$ and $P_\f$, and the corresponding Gaussian integrals can be carried out to get the Lagrangian path integral with a nontrivial field dependent measure. 

Having obtained the complete gauge fixed action, one may look for residual symmetries which are possibly left over. It is easy to see that \eq{23} completely eliminates diffeomorphisms with $k^0\not=0$. To proceed, it is convenient to introduce a background value for the trace of $h_{ij}$  and write
\be\label{33}
h_{ij}=a(t)^2(1+\z)\d_{ij}+\cc_{ij}
\ee
where $a(t)$ is (for now) an arbitrary non-vanishing function of time. Under a possibly time dependent map generated by $k^i(t,\vec{x})$, the tensor $\cc_{ij}$ transforms as
\be
\d_S\cc_{ij}=k^r\del_r\cc_{ij}+\cc_{rj}\del_ik^r+\cc_{ir}\del_jk^r-\fr23\d_{ij}\cc_{rs}\del_sk^r
+a^2(1+\z)\left(\del_jk^i+\del_ik^j-\fr23\d_{ij}\del_rk^r\right).\label{34}
\ee
The residual diffeomorphisms must satisfy
\be
\del_i(\d_S\cc_{ij})=0,\label{35}
\ee
which gives
\be
\del^2k^j+\fr13 \del_i\del_jk^i=. . . \la{36}
\ee
where $\del^2=\del_i\del_i$ and the right hand side contains $\z$ and $\cc_{ij}$ dependent terms. It is easy to see that the dilatation, which is generated by $k^i=x^i$, exactly solves \eq{36} where we refer to the ADM coordinates introduced in \eq{13}. There are also field dependent solutions that can be expressed as a series in the field variables $\z$ and $\cc_{ij}$ where the zeroth order term solves the left hand side of \eq{36}. In principle, this yields an infinite set of residual diffeomorphisms for each zeroth order solution \cite{in1,in2}, but it is very unlikely that these survive as the symmetries of the quantum theory. It is easy to see that a field dependent map does not generate a canonical transformation; in general the Poisson bracket structure is destroyed
\be\la{37}
\d_S\left\{h_{ij},\pi^{rs}\right\}\not=0.
\ee
This can also be verified by observing that the canonical form of the action $\int d^4x\,\Pi^{ij}\dot{h}_{ij}$ is not going to be preserved after the transformation. Therefore, the path integral measure picks up a nontrivial Jacobian which would ruin the symmetry. Moreover, the non-covariant ghost action contains usual partial derivatives and in general it will not be invariant under a field dependent diffeomorphism. 

This leaves the dilatation as the only residual symmetry candidate and one must still check the invariance of the total action. The variation of the Einstein-Hilbert action gives a surface term as indicated in \eq{18}. One may observe that the gauge conditions \eq{23} ad \eq{25} transform like tensor densities 
\bea
&&\d_DG=x^i\del_iG,\nn\\  
&&\d_DG_i=x^j\del_jG_i+3G_i.\label{38}
\eea
Therefore, the gauge fixing action \eq{26}  changes up to surface terms under the dilatation if one imposes the Lagrange multipliers to obey
\bea
&&\d_D\l=x^i\del_i\l+3\l,\nn\\
&&\d_D\l_i=x^j\del_j\l_i.\la{39}
\eea
The nice tensor-type transformation properties of the fields under the dilatation also implies the invariance of the ghost action as follows: By definition, the constraints $\F$ and $\F_i$ are tensor densities. It is an elementary exercise to show that under an infinitesimal canonical transformation the variation of a Poisson bracket is equal to the Poisson bracket of the variation, i.e.
\be\la{40}
\d\{A,B\}=\{\d A,B\}+\{A,\d B\},
\ee 
which shows that the Faddeev-Popov matrix entries $\{\F_\m,G_\n\}$ transforms like bi-tensor densities under the dilation. From the corresponding wights, the transformation of the ghost fields that would leave the ghost action invariant (up to surface terms) can be found as
\bea
&&\d_D\chi=x^j\del_j\chi+3\chi,\nn\\
&&\d_D\chi^i=x^j\del_j\chi^i,\nn\\
&&\d_D\cb=x^j\del_j\cb,\la{41}\\
&&\d_D\cb^i=x^j\del_j\cb^i-\cb^i.\nn
\eea
As a result, the variation of the complete action $S=S_{EH}+S_{GF}+S_{FP}$ under the dilatation $k^i=x^i$ becomes a surface term, where the fields are mapped as in \eq{17}, \eq{39} and \eq{41}. The path integral measure only picks up an irrelevant constant Jacobian under these transformations, which act like a canonical map\footnote{Although the deformed transformation is not canonical, it is nevertheless linear and thus the phase space path integral measure at most picks up an irrelevant constant as in the case of Lagrange multipliers and ghosts.} for the fields $h_{ij}$, $\Pi^{ij}$, $\f$, $P_\f$, and as a linear map for the ghosts $\chi^\m$, anti-ghosts $\cb^\m$ and the Lagrange multipliers $\l^\m$. So the dilatation becomes the symmetry of the quantum theory when suitable boundary conditions killing the surface terms are imposed. 

\section{Perturbative Expansion, Boundary Conditions and the Noether Charge}

In this section we quantify our previous findings by an expansion around a cosmological FRW background $ds^2=-dt^2+a^2dx^idx^i$. We introduce the fluctuation fields $(\z,P_\z)$, $(\cc_{ij},\C^{ij})$, $(\vf,P_\vf)$, $n$ and $n^i$ as
\bea
&&h_{ij}=a^2(1+\z)\d_{ij}+\cc_{ij},\nn\\
&&\Pi^{ij}=\left(-2\dot{a}+P_\z/2a^2\right)\d_{ij}+\C^{ij},\nn\\
&&\f=\f_B+\vf,\nn\\
&&P_\f=a^3\dot{\f}_B+P_\vf,\la{42}\\
&&N=1+n,\nn\\
&&N^i=n^i,\nn
\eea
where $\d_{ij}\C^{ij}=0$, and assume that the following background equations are satisfied
\be
6H^2=\fr{1}{2}\dot{\f}_B^2+ V_B,\hs{10} \dot{H}=-\fr14\dot{\f}_B^2,\la{43}
\ee
where $H=\dot{a}/a$ and $V_B=V(\f_B)$. The ghost fields and the Lagrange multipliers have no background values and the pairs $(\z,P_\z)$, $(\cc_{ij},\C^{ij})$, $(\vf,P_\vf)$ are canonical conjugates. After integrating out $\l^\m$, which enforces the gauge conditions $\vf=0$ and $\del_i\cc_{ij}=0$, the path integral measure becomes
\be
DX\equiv D\z DP_\z D^T\cc_{ij}D\C^{kl}Dn Dn^i  D\cb^\m D\chi^\n\la{44}
\ee
where $D^T\cc_{ij}$ denotes the sum over only the transverse tensor modes. Evidently, in expanding the action around the background solution inside the path integral, both conditions $\vf=0$ and $\del_i\cc_{ij}=0$ can freely be used. A straightforward calculation then yields
\bea
S=&&\int\C^{ij}\dot{\cc}_{ij}+P_\z\dot{\z}+\fr{1}{6a^3}P_\z^2+H\z P_\z-\fr{1}{2a^3}P_\vf^2-a\C^{ij}\C^{ij}+2H\C^{ij}\cc_{ij}+\fr32\dot{\f}_B\z P_\vf\nn\\
&&+(\dot{H}-H^2)\cc_{ij}\cc_{ij}-\left(3a^2\ddot{a}+\fr32 a^3\dot{\f}_B^2\right)\z^2+\fr12a\del_k\z\del_k\z-\fr{1}{4a^3}\del_k\cc_{ij}\del_k\cc_{ij}+. . .\la{45}\\
&&-n\tilde{\F}-n^i\tilde{\F}_i\nn
\eea
where
\bea
&&\tilde{\F}=\dot{\f}_BP_\vf+2a\del^2\z+2HP_\z+6a^2\ddot{a}\z+. . . \nn\\
&&\tilde{\F}_i=-2a^2\del_j\C^{ij}-\fr23\del_iP_\z-2a^2\dot{a}\del_i\z+. . .\la{46}
\eea
and only the quadratic fluctuation terms in the action are explicitly found whereas the higher order terms are simply indicated by dots. Since we are expanding around a classical solution, the linear terms in the action cancel each other. We omit writing the ghost action since the ghosts decouple from the fluctuations at the quadratic order. As a check of \eq{45}, one may solve the constraints in \eq{46} as
\bea
&&P_\vf=-\fr{2a}{\dot{\f}_B}\del^2\z-\fr{2\dot{a}}{a\dot{\f}_B}P_\z-\fr{6a^2\ddot{a}}{\dot{\f}_B}\z,\nn\\
&&\C^{ij}=\C^{ij}_T+\fr{1}{6a^2}\d_{ij}P_\z+\fr{\dot{a}}{2}\d_{ij}\z-\fr{1}{2a^2}\del_i\del_j\fr{1}{\del^2}P_\z-\fr{3\dot{a}}{2}\del_i\del_j\fr{1}{\del^2}\z,\la{47}
\eea
where $\del_i\C_T^{ij}=0$ and $\C_T^{ii}=0$. Using these solutions back in the action \eq{45} and after eliminating $P_\z$ through its equation of motion gives the quadratic action
\be\label{48}
S^{(2)}=\int \fr12a^3\e\dot{\z}^2-\fr12a\e\del_i\z\del_i\z+\fr{1}{4a}\dot{\cc}_{ij}\dot{\cc}_{ij}-\fr{1}{4a^3}\del_k\cc_{ij}\del_k\cc_{ij}+\fr{\ddot{a}}{2a^2}\cc_{ij}\cc_{ij},
\ee
where $\e=\dot{\f}_B^2/4H^2=-\dot{H}/H^2$ is the slow-roll parameter. This becomes the standard quadratic action of cosmological perturbations after rescaling $\cc_{ij}\to a^2\cc_{ij}$, which transforms our $\cc_{ij}$ to the linearized graviton field used in the literature.

As shown in the previous section, the nonlinear action including  ghosts is invariant under the dilatation. While the variation of the ghosts are still given by \eq{41}, the fluctuation fields can be found to transform as
\bea
&&\d_D P_\vf=x^i\del_i P_\vf+3P_\vf+3a^3\dot{\f}_B,\nn\\
&&\d_D\z=x^i\del_i\z+2\z+2,\nn\\
&&\d_D P_\z=x^i\del_iP_\z+P_\z-6a^2\dot{a},\nn\\
&&\d_D \cc_{ij}=x^k\del_k\cc_{ij}+2\cc_{ij},\la{49}\\
&&\d_D\C^{ij}=x^k\del_k\C^{ij}+\C^{ij},\nn\\
&&\d_D n=x^i\del_i n,\nn\\
&&\d_D n^i=x^k\del_k n^i-n^i.\nn
\eea 
The change of the action under the deformed dilatation $\d_\r=\r\,\d_D$ becomes
\be
\d_\r S=-\int d^4 x\,\r\,[\del_\m J^\m],\la{50}
\ee
where the corresponding Noether current (below $O(2)$ denotes the quadratic and the higher order fluctuation terms) is given by
\bea
&&J^0=2P_\z+6a^3H\z+O(2),\nn\\
&&J^i=2a\del_i\z+4a\del_in+O(2).\la{51}
\eea
We check $\del_\m J^\m=0$ provided that the linearized field equations are obeyed. We also confirm that to that order $\del_\m J^\m=\del_\m\hat{J}^\m$, where $\hat{J}^\m$ is the full nonlinear current given in \eq{21}. 
 
Like \eq{22}, the charge density can be expressible as
\be\label{52}
J^0=x^i\tilde{\F}_i+\del_i\left[x^i\left(\fr23 P_\z+2a^2\dot{a}\z\right)+2a^2x^j\C^{ij}+O(2)\right].
\ee
Thus, similar to the full nonlinear charge density \eq{22}, its perturbative version \eq{52} can also be written as the sum of the momentum constraint and a total divergence term. Indeed, defining a new Noether current using the freedom \eq{fr} as
\bea
&&\tilde{J}^0=\hat{J}^0+\del_i\left[6a^2\dot{a}x^i\z+4a^2\dot{a}x^i\right],\nn\\
&&\tilde{J}^i=\hat{J}^i-\fr{d}{dt}\left[6a^2\dot{a}x^i\z+4a^2\dot{a}x^i\right], \la{53}
\eea
one may see that $\tilde{J}^0$ and $J^0$ agree on linear terms, which prove \eq{52} (one may replace $J^\m$ with $\tilde{J}^\m$ if necessary). On shell, i.e. when the momentum constraint is obeyed, the Noether charge becomes
\be\la{54}
Q=\int d^3x\,J^0=\lim_{r\to\infty}r^3\int_{S^2} d\O\left[\fr23 P_\z+2a^2\dot{a}\z+2a^2\fr{x^ix^j}{r^2}\C^{ij}+O(2)\right],
\ee
which is a surface integral at spatial infinity. 

The form of the surface charge \eq{54} suggests the following boundary conditions for the fields
\be\la{55}
\z,\,P_\z,\,\cc_{ij},\,\C^{ij}=O\left(\fr{1}{r^3}\right)\hs{5}\textrm{as}\hs{5}r\to\infty.
\ee
We assume that the time derivatives of the fields have the same fall-off rates and a spatial derivative increases the order by one like $\del_i\z=O(1/r^4)$.  Together with these boundary conditions the operator $1/\del^2$ becomes well defined in the position space, for example one has
\be\la{56}
\fr{1}{\del^2}\z(\vec{x})=-\fr{1}{4\pi}\int d^3y\fr{\z(\vec{y})}{|\vec{x}-\vec{y}|}.
\ee
Therefore $(1/\del^2) \z=O(1/r)$ and the standard solutions of the lapse $n$ and the shift $n^i$  in the linearized theory suggests
\be\la{57}
 n=O\left(\fr{1}{r^3}\right),\hs{5}n^i=O\left(\fr{1}{r^2}\right)\hs{5}\textrm{as}\hs{5}r\to\infty.
\ee
The fall-off conditions \eq{55} and \eq{57} apply to all fields in the path integral. Moreover, from $J^\m$ given in \eq{51} one sees that the current conservation $\del_\m J^\m=0$ implies 
\be\la{58}
\dot{Q}=0,
\ee
i.e. the spatial flux of the field at infinity vanishes and the Noether charge is conserved. 

Consider now the expectation value of an operator $O(t)$, conveniently defined at a single time, given by the in-in path integral. From the infinitesimal variations of the integration variables, one may get the Schwinger-Dyson equations $\lf [E.o.M.] O(t)\rg=0$ and $\lf O(t)[E.o.M.] \rg=0$. Specifically, taking the operator in the plus-branch $O_+(t)$ in the path integral and varying $n^i_-$ and choosing $O_-(t)$ in the path integral and varying $n^i_+$ respectively give
\be\la{59}
\lf O(t)\F_i\rg=\lf \F_i O(t)\rg=0.
\ee
Furthermore, manipulations similar to the previous section imply
\be\la{60}
\lf O(t)[\del_\m J^\m] \rg=\lf [\del_\m J^\m] O(t)\rg=0
\ee
and 
\bea
&&\lf \dot{Q}O(t)\rg=\lf O(t)\dot{Q}\rg=0,\nn\\
&&i\lf\left[Q,O(t)\right]\rg=\lf\d_D O(t)\rg.\la{61}
\eea
The prescribed boundary conditions are only used in the last equation to make sure the existence of the Noether charge in the path integral. We note that these identities are valid for any initial state in the theory.

The Noether charge inside the path integral can be expressed as a surface charge depending on fields at infinity since the momentum constraint is satisfied by the Schwinger-Dyson equations \eq{59}. In integrating out the boundary fields living at spatial infinity, the full action can be replaced by the quadratic one because of the presumed boundary conditions (in other words, the full interacting theory becomes linearized and free near spatial infinity). Using the linearized momentum constraint \eq{46}, the surface integral can be converted back to a volume integral so that
\be
Q=\int d^3 x\left[2P_\z^{(f)}+6a^2\dot{a}\z^{(f)}\right],\la{62}
\ee
where the label $(f)$ signifies that these are free fields governed by the quadratic action. One may further use the linearized equation $P_\z^{(f)}=a^3\e \dot{\z}^{(f)}-3a^2\dot{a}\z^{(f)}-(a^2/\dot{a})\del^2\z^{(f)}$ to obtain
\be\la{63}
Q=2\int d^3 x\, \left[a^2\,\e\,\dot{\z}^{(f)}\right].
\ee
From the field equation of $\z^{(f)}$ and using the fall-off conditions, one may readily verify that $\dot{Q}=0$. 

Formally, it is possible to obtain \eq{63} in the operator formalism as follows: Take the variable $\z$. The exact Heisenberg picture operator $\z_H$ is related to the free interaction picture operator $\z^{(f)}$ by
\be\la{64}
\z_H(t,\vec{x})=\z^{\I}(t,\vec{x})-i\int_{t_i}^t dt'\left[\z^{\I}(t,\vec{x}),H_I(t')\right]+. . . 
\ee
where $H_I(t')=\int d^3 y {\cal H}_I(t',\vec{y})$ is the interaction Hamiltonian and the dotted terms contain more nested commutators of $\z^{\I}$ with $H_I$. By causality, for any given $\vec{y}$ one has
\be\la{65}
\lim_{|\vec{x}|\to\infty}\left[\z^{\I}(t,\vec{x}),{\cal H}_I(t',\vec{y})\right]= 0,
\ee
therefore one expects 
\be\la{66}
\lim_{|\vec{x}|\to\infty}\left[\z^{\I}(t,\vec{x}),H_I(t')\right]\to 0,
\ee
which would imply 
\be\la{67}
\lim_{|\vec{x}|\to\infty}\z_H(t,\vec{x})\to\lim_{|\vec{x}|\to\infty}\z^{\I}(t,\vec{x}).
\ee
However, the Hamiltonian $H_I(t')$ is given by the integral of the Hamiltonian density ${\cal H}_I(t',\vec{y})$ extending to spatial infinity, so $\z^{\I}(t,\vec{x})$ and $H_I(t')$ have causally overlapping regions even when $|\vec{x}|\to\infty$. Consider a similar commutator
\be\la{68}
\left[\z^{\I}(t,\vec{x}),\tilde{\z}^{\I}(t',\vec{k})\right]\propto e^{-i\vec{k}.\vec{x}}
\ee
where $\tilde{\z}^{\I}$ is the Fourier transform of $\z^{\I}$ defined by
\be\la{69}
\tilde{\z}^{\I}(t',\vec{k})=(2\pi)^{-3/2}\int d^3 x\, e^{-i\vec{k}.\vec{x}}\,\z^{\I}(t',\vec{x}),
\ee
which involves an integral extending to spatial infinity as in the interaction Hamiltonian. Although \eq{68} does not converge  to zero point-wise as $|\vec{x}|\to\infty$, its angular average in the position space produces the factor $\sin(k|\vec{x}|)/(k|\vec{x}|)$ that vanishes in the limit of interest. By expressing the interaction Hamiltonian in terms of the Fourier transformed variables, one may then predict that the commutator \eq{66} vanishes after the angular integration is carried out as $|\vec{x}| \to0$ implying 
\be
\lim_{r\to\infty}\int_{S^2}d\O\,\z_H(t,r,\th,\f)\to\lim_{r\to\infty}\int_{S^2}d\O\,\z^{\I}(t,r,\th,\f).\la{70}
\ee
This shows that the Noether charge operator given in \eq{54} can be expressed by the free fields, which straightforwardly leads to \eq{63}. 

Using the free field expansion
\be\la{71}
\z^{(f)}=(2\pi)^{-3/2} \int d^3k\,e^{i\vec{k}.\vec{x}}\,\m_k(t)\,a_{\vec{k}}+h.c.
\ee
where
\be
\m_k\dot{\m}_k^*-\m_k^*\dot{\m}_k=\fr{i}{\e a^3},\la{72}
\ee
one may find
\be\la{73} 
\left.Q\vv=\lim_{k\to0} 2(2\pi)^{3/2}a^3\e\left(\fr{\dot{\m}_k}{\m_k}\right)^*\tilde{\z}^{(f)}(\vec{k})\left. \vv,
\ee
where $a_{\vec{k}}\left.\vv=0$. The standard curvature perturbation $\hat{\z}$, which is conserved at the superhorizon scales, is defined by
\be\la{74}
\hat{\z}=\fr16\ln\left[\fr{1}{a^3}\det h_{ij}\right].
\ee
At the linearized level, $\hat{\z}$ and $\z$ agree with each other. Since the zero mode of $\hat{\z}$ is constant and the Heisenberg picture operators are identified with the corresponding interaction picture operators at $t_i$,  one has\footnote{For notational simplicity the Fourier transform of the operator $\hat{\z}(t,\vec{x})$, which is defined as in \eq{69}, is denoted by the same symbol  $\hat{\z}(t,\vec{k})$. Since $\lim_{k\to0}\hat{\z}(t,\vec{k})$ is conserved its time argument is also dropped.} $\lim_{k\to0}\hat{\z}(\vec{k})=\lim_{k\to0}\tilde{\z}^{(f)}(\vec{k})$ giving 
\be\la{75} 
\left.Q\vv=\lim_{k\to0} 2(2\pi)^{3/2}a^3\e\left(\fr{\dot{\m}_k}{\m_k}\right)^*\hat{\z}^{(f)}(\vec{k})\left. \vv,
\ee
Then, for any operator that commutes with $\hat{\z}(\vec{k}=0)$ like $O(t)=\hat{\z}(t,\vec{x}_1). . .  \hat{\z}(t,\vec{x}_n)$, the equations \eq{72} and \eq{75} yield
\be\la{76}
\lim_{k\to0}\fr{2(2\pi)^{3/2}}{|\m_k|^2}\lf O(t)\hat{\z}(\vec{k})\rg=\lf\d_D O(t)\rg,
\ee
which is the dilatational consistency relation of cosmological perturbations. As an example, by taking $O(t)=\hat{\z}(t,\vec{k}_1)\hat{\z}(t,\vec{k}_2)$ in \eq{76} and noting that $\d_D\hat{\z}(t,\vec{x})=x^i\del_i\hat{\z}+1$, one may obtain
\be\la{77}
\lim_{k\to0}\fr{2(2\pi)^{3/2}}{P(k)}\lf \hat{\z}(t,\vec{k}_1)\hat{\z}(t,\vec{k}_2)\hat{\z}(\vec{k})\rg=-\d^3(\vec{k}_1+\vec{k}_2)\del_i^{(k_1)}\left[k^i_1P(k_1)\right],
\ee
where $P(k)$ is the exact two point function in the momentum space, which is defined by $\lf\hat{\z}(t,\vec{k}_1)\hat{\z}(t,\vec{k}_2)\rg=\d^3 (\vec{k}_1+\vec{k}_2)P(k)$ and satisfies $\lim_{k\to0}P(k)=\lim_{k\to0}|\m_k|^2$. Eq. \eq{77} can be identified as the consistency relation for the 3-point function \cite{mal,cr1}. 

\section{Conclusions}

In this paper we try to clarify a few issues about the cosmological Ward identities related to the residual symmetries of the cosmological perturbations in the phase space path integral quantization method. The general role played by the Noether current in the path integral approach is reviewed. We study the gravity plus minimally coupled self-interacting scalar field system in the so called $\z$-gauge, which is relevant for the slow-roll inflation. The ghost action yielding the Faddeev-Popov determinant is obtained. We observe that only the dilatation survives as the residual symmetry of the complete gauge fixed action including the ghosts. The corresponding Noether current is calculated both exactly in the nonlinear theory and in the linearized form when the theory is expanded around a cosmological background. The Noether charge is shown to be equivalent to a surface integral at spatial infinity and the boundary conditions that are required for charge conservation are identified. It turns out that the charge can be related to the zero mode of the curvature perturbation $\z$ and the related Ward identity gives the consistency relation of the cosmological perturbations. 

The present work can be extended in a few directions. Since the full Noether current is obtained, it is possible to determine the nonzero momentum corrections to the cosmological consistency relation in a systematic way (see \cite{m1,m2,m3}). Indeed, it is not difficult to get the nonzero momentum version of the identity \eq{61} that involves the Noether charge carrying momentum $\vec{k}$ defined by $Q(\vec{k})=\int d^3x \exp(i\vec{k}.\vec{x})J^0(t,\vec{x})$. 
Introducing $J^i(\vec{k})=\int d^3x \exp(i\vec{k}.\vec{x})J^i(t,\vec{x})$, the current conservation implies $\dot{Q}(\vec{k})=ik_iJ^i(\vec{k})$, thus one would expect the vector $J^i$ to show up in the Ward identity. This is work in progress which we hope to report soon. More ambitiously, it would be interesting to generalize the present formalism to accommodate other field dependent residual symmetries discussed in the literature. The main obstacle here is that a field dependent diffeomorphism do not leave the path integral measure invariant. One plausible way of avoiding this issue is to focus on the asymptotic symmetries as in \cite{as1,as2,as3,as4,as5,as6}. It might also be possible to find nicer field variables or alternative gauge conditions allowing different residual symmetries. 

In the context of gauge-gravity duality, the asymptotic charges are usually defined using the counter-term subtraction method and in \cite{son1} these have been shown to generate the desired asymptotic symmetries of the AdS space. On the other hand, in \cite{son2} the asymptotic symmetry group and the corresponding charges are specified on appropriately constructed phase space for the asymptotically de Sitter Einstein gravity. Curiously, the spatial fall-off conditions utilized in \cite{son2} for the construction of the phase space are very similar to the ones imposed in this paper. It would be interesting to study both of these approaches in the context of scalar inflationary models and examine their implications for the cosmological correlation functions. These are important questions that are worth to dwell on, whose answers are expected to improve our understanding of the cosmological perturbations at the nonlinear quantum level. 

\begin{acknowledgments}
This work, which has been done at C.\.{I}.K. Ferizli, Sakarya, Turkey without any possibility of using references, is dedicated to my friends at rooms C-1 and E-10 who made my stay bearable at hell for 440 days between 7.10.2016 and 20.12.2017. I am also indebted to the colleagues who show support in these difficult times. 
\end{acknowledgments}

\end{document}